\documentstyle[12pt,epsf,psfig,fleqn]{article}
\DeclareFontFamily{U}{msb}{}
\DeclareFontShape{U}{msb}{m}{n}{
<5><6><7><8><9> gen *msbm <10><10.95><12><14.4><17.28><20.74><24.88>msbm10}{}
\DeclareSymbolFont{AMSb}{U}{msb}{m}{n}
\DeclareMathSymbol{\AAA}{\mathbin}{AMSb}{'101}
\DeclareMathSymbol{\BBB}{\mathbin}{AMSb}{'102}
\DeclareMathSymbol{\CCC}{\mathbin}{AMSb}{'103}
\DeclareMathSymbol{\DDDD}{\mathbin}{AMSb}{'104}
\DeclareMathSymbol{\EEE}{\mathbin}{AMSb}{'105}
\DeclareMathSymbol{\FFF}{\mathbin}{AMSb}{'106}
\DeclareMathSymbol{\GGG}{\mathbin}{AMSb}{'107}
\DeclareMathSymbol{\HHH}{\mathbin}{AMSb}{'110}
\DeclareMathSymbol{\III}{\mathbin}{AMSb}{'111}
\DeclareMathSymbol{\JJJ}{\mathbin}{AMSb}{'112}
\DeclareMathSymbol{\KKK}{\mathbin}{AMSb}{'113}
\DeclareMathSymbol{\LLL}{\mathbin}{AMSb}{'114}
\DeclareMathSymbol{\MMM}{\mathbin}{AMSb}{'115}
\DeclareMathSymbol{\NNN}{\mathbin}{AMSb}{'116}
\DeclareMathSymbol{\OOO}{\mathbin}{AMSb}{'117}
\DeclareMathSymbol{\PPPP}{\mathbin}{AMSb}{'120}
\DeclareMathSymbol{\QQQ}{\mathbin}{AMSb}{'121}
\DeclareMathSymbol{\RRR}{\mathbin}{AMSb}{'122}
\DeclareMathSymbol{\SSS}{\mathbin}{AMSb}{'123}
\DeclareMathSymbol{\TTTTT}{\mathbin}{AMSb}{'124}
\DeclareMathSymbol{\UUU}{\mathbin}{AMSb}{'125}
\DeclareMathSymbol{\VVVV}{\mathbin}{AMSb}{'126}
\DeclareMathSymbol{\WWW}{\mathbin}{AMSb}{'127}
\DeclareMathSymbol{\XXXX}{\mathbin}{AMSb}{'130}
\DeclareMathSymbol{\YYY}{\mathbin}{AMSb}{'121}
\DeclareMathSymbol{\ZZZZ}{\mathbin}{AMSb}{'132}

\newcommand{\Int}{\int\limits}
\newcommand{\vv}{\alpha}
\newcommand{\rr}{\vec{r}}
\newcommand{\qq}{\vec{q}}
\newcommand{\Fa}{\mbox{\eusxii F}}
\newcommand{\La}{\mbox{\eusxii L}}

\newcommand{\rao}{\rightarrow 0}

\newcommand{\rai}{\rightarrow \infty}

\newcommand{\al}{\alpha}

\newcommand{\DIF}[2]{\partial^{#1}_{#2}}

\begin{document}
\newfont{\eufxii}{eufm10 scaled 1200}
\newfont{\eusix}{eusm10 scaled 900}
\newfont{\eusxii}{eusm10 scaled 1200}
\newfont{\eurxii}{eurm10 scaled 1200}
\setcounter{page}{0}
\begin{center}
{\Large\bf On Fractional Diffusion and its Relation with\\[12pt]
Continuous Time Random Walks}\\[24pt]
{R. Hilfe$\mbox{\rm r}^{1,2}$}\\[24pt]
{\em 
$\mbox{ }^1$ICA-1, Universit{\"a}t Stuttgart,
Pfaffenwaldring 27, 70569 Stuttgart\\
$\mbox{ }^2$Institut f{\"u}r Physik,
Universit{\"a}t Mainz,
55099 Mainz, Germany}\\[30pt]
\thispagestyle{empty}
ABSTRACT
\end{center}
Time evolutions whose infinitesimal generator is a 
fractional time derivative arise generally in the
long time limit.
Such fractional time evolutions are considered here
for random walks.
An exact relationship is established between the
fractional master equation and a separable continuous 
time random walk of the Montroll-Weiss type.
The waiting time density can be expressed using a generalized 
Mittag-Leffler function.
The first moment of the waiting density does not exist.\\[3cm]
in: \\
{\em Anomalous Diffusion - From Basics to Applications},
R. Kutner, A. Pekalski and K. Sznaij-Weron (eds.),
Lecture Notes in Physics, vol. 519,
Springer, Berlin 1999, pages 77-82
\newpage
\section{Fractional Time Evolution}
A series of recent investigations \cite{hil93e,hil95c,hil95e,hil95f,hil98c}
has found that, in a suitable long time limit, the macroscopic
time evolution $T^t$ of a physical observable $X(t)$ is 
given as a convolution of the form 
\begin{equation}
T^t_\al X(t_0)
= \Int_0^\infty X(t_0-s)
h_\al\left(\frac{s}{t}\right)\frac{ds}{t}
\label{globevol}
\end{equation}
where $t\geq 0$, $0<\al\leq 1$ and
\begin{equation}
h_\al(x) = \frac{1}{x\al}
H^{10}_{11} \left(\frac{1}{x} \left|
\begin{array}{cc}
(0,1)\\
(0,1/\al) 		
\end{array}
\right. \right)
\label{Hfunc}
\end{equation}
is defined through its Mellin transform \cite{fox61}
\begin{equation}
\int_0^\infty H^{10}_{11} \left(x \left|
\begin{array}{cc}
(0,1)\\
(0,1/\al)
\end{array}
\right. \right) 
x^{s-1} \; dx =
\frac{\Gamma(s/\al)}{\Gamma(s)} 
\end{equation}
For $\al=1$ 
\begin{equation}
h_1(x) = \lim_{\al\rightarrow1^-}h_\al(x) = \delta(x-1)
\label{e460}
\end{equation}
and hence eq. (\ref{globevol}) reduces to
\begin{equation}
T^t_1X(t_0)=X(t_0-t)
\end{equation}
a simple translation.
For fixed $\al$ the operators $T^t_\al,t\geq 0$ form a one-parameter 
semigroup on a Banach space (e.g. the space of all continuous functions 
vanishing at infinity).

Much interest in the semigroups $T^t_\al$ derives from the
fact that their infinitesimal generators are proprotional to
\begin{equation}
(\DIF{\al}{+}X)(t)=\lim_{s\rao}\frac{T^s_\al X(t)-X(t)}{s} = 
\frac{1}{\Gamma(-\al)}\Int_0^\infty \frac{X(t)-X(t-s)}{s^{\al+1}}ds ,
\end{equation}
i.e. to fractional time derivatives of order $\al$ \cite{hil95f}.
This may be seen from Ref.\cite{fel71} (p. 302), and the fact that the
kernels $h_\al(s/t)/t$ in eq. (\ref{globevol}) are stable
probability densities (see also \cite{hil98c}).

Derivatives of noninteger order $0<\al\leq 1$ with respect to time 
are in this way found to furnish an important generalization for the 
infinitesimal generator of the macroscopic time evolution of physical 
observables.

Given the generality of the results above it is natural
to consider specific examples by replacing the integer
order time derivative with a fractional derivative.
A particularly interesting example is given by diffusion
and random walks \cite{hil95a,hil95b}.
The fractional generalization of the familiar master
equation for random walks turns out to be intimately 
related to the theory of continuous time random walks 
with power-law tails in the waiting time density
\cite{KBS87}.
It is the purpose of this note to exhibit an exact 
relationship between a fractional master equation and
continuous time random walks \cite{hil95a}.

Let me conclude this introduction with the remark
that only a special subclass of separable 
continuous time random walks with algebraically decaying 
waiting time distributions corresponds exactly to a 
generalized master equation with Riemann-Liouville-type
fractional time derivative.
It would be interesting to specify precisely whether,
and under which conditions, other waiting time densities
with a power-law tail lead to the same result \cite{com96,MKS98}.

\section{Fractional Master Equation}

Consider a random walk and
let $p(\rr,t)$ denote the probability density to find
the walker at the position $\rr\in\RRR^d$ at time $t$
if it was at the origin $\rr=0$ at time $t=0$. 
The positions
$\rr\in\RRR^d$ may be discrete or continuous.
The conventional master equation for the random walk
reads
\begin{equation}
\frac{\partial}{\partial t}
p(\rr,t) = \sum_{\rr^\prime}w(\rr-\rr^\prime)p(\rr^\prime,t)
\label{me}
\end{equation}
with initial condition $p(\rr,0) = \delta_{\rr0}$. 
To establish the fractional generalization of this initial
value problem one must respect the nonlocal nature of the
fractional derivatives.
This leads to the fractional master equation
\begin{equation}
\DIF{\vv}{0+}p(\rr,t) = \delta_{\rr0}\frac{1}{\Gamma(1-\vv)t^{\vv}} + 
\sum_{\rr^\prime}w(\rr-\rr^\prime)p(\rr^\prime,t)
\label{e110}
\end{equation}
where $0<\vv\leq 1$, $t\geq 0$, and the initial condition 
$p(\rr,0)=\delta_{\rr0}$ has been incorporated.
Note that the fractional transition rates $w(\rr)$ 
have units of (1/time$)^\vv$.
They obey the relation $\sum_{\rr}w(\rr)=0$. 
Applying a fractional integral of order $\vv$ to equation 
(\ref{e110}) yields an integral equation
\begin{equation}
p(\rr,t)= \delta_{\rr0}+\frac{1}{\Gamma(\vv)}
\int_0^t(t-t^\prime)^{\vv-1}\sum_{\rr^\prime}
w(\rr-\rr^\prime)p(\rr^\prime,t^\prime)\,dt^\prime
\label{e120}
\end{equation}
reminiscent of the integral equation for continuous
time random walks \cite{BN70,MW79,WR83,HK87,hug95,hug96}.

\section{Continuous Time Random Walks}

The basic integral equation for separable continous time random walks
describes a random walker in continous time without correlation
between its spatial and temporal behaviour, and reads \cite{HK87,KBS87}
\begin{equation}
p(\rr,t) = \delta_{\rr 0} \Phi(t) + \int_0^t \psi(t-t^\prime)
\sum_{\rr^\prime}
\lambda(\rr-\rr^\prime)p(\rr^\prime,t^\prime)\,dt^\prime .
\label{e210}
\end{equation}
Here, as in (\ref{e120}),
$p(\rr,t)$ denotes the probability density to find the walker
at the position $\rr\in\RRR^d$ at time $t$ if it started from 
the origin $\rr=0$ at time $t=0$. 
$\lambda(\rr)$ is the probability 
for a displacement $\rr$ in each single step, and $\psi(t)$ is the
waiting time distribution giving the probability density
for the occurrence of a time interval $t$ between two consecutive steps. 
The transition probabilities obey $\sum_{\rr}\lambda(\rr)=1$. 
The function $\Phi(t)$ in eq. (\ref{e210}) 
is the survival probability at the initial 
position.
It is related to the waiting time distribution through
\begin{equation}
\Phi(t) = 1 - \int_0^t\psi(t^\prime)\,dt^\prime.
\label{e220}
\end{equation}
Note that in equation (\ref{e210}) the system is prepared
with the walker at position $\rr=0$, and it develops from
there according to $\psi(t)$.

\section{Relation between equations (\ref{e120}) and (\ref{e210})}

The similarity between equations (\ref{e120}) and (\ref{e210})
suggests that the former is a special case of the latter.
To show that this is true let
\begin{equation}
\psi(u)=\La\{\psi(t)\}(u)=\int_0^\infty e^{-ut}\psi(t)\,dt
\label{e230}
\end{equation}
denote the Laplace transform of $\psi(t)$ and write
\begin{equation}
\lambda(\qq)=\Fa\{\lambda(\rr)\}(\qq)=\sum_{\rr}e^{i\qq\cdot\rr}\lambda(\rr)
\label{e240}
\end{equation}
for the Fourier transform of $\lambda(\rr)$.
Then the Fourier-Laplace transform $p(\qq,u)$ of the 
solution to (\ref{e210}) is given as \cite{BN70,MW79,WR83,KBS87}
\begin{equation}
p(\qq,u) = \frac{1}{u}\frac{1-\psi(u)}{1-\psi(u)\lambda(k)} = 
\frac{\Phi(u)}{1-\psi(u)\lambda(\qq)}
\label{e250}
\end{equation}
where $\Phi(u)$ is the Laplace transform of the survival probability.

Similarly the fractional master equation (\ref{e120}) can be solved
in Fourier-Laplace space with the result
\begin{equation}
p(\qq,u) = \frac{u^{\vv-1}}{u^\vv-w(\qq)}
\label{e260}
\end{equation}
where $w(\qq)$ is the Fourier transform of the kernel $w(\rr)$ 
in(\ref{e120}). Eliminating $p(\qq,u)$ between (\ref{e250}) and
(\ref{e260}) gives the result
\begin{equation}
\frac{1-\psi(u)}{u^\vv\psi(u)} = \frac{\lambda(\qq)-1}{w(\qq)} = C
\label{e270}
\end{equation}
where $C$ is a constant. The last equality holds because the 
left hand side of the first equality is $\qq$-independent while
the right hand side is independent of $u$. 

From (\ref{e270}) it is seen that the fractional 
master equation characterized by the kernel $w(\rr)$ and the
order $\vv$ corresponds to a special class of space time 
decoupled continuous time random walks characterized by 
$\lambda(\rr)$ and $\psi(t)$. 
This correspondence is given precisely as
\begin{equation}
\psi(u) = \frac{1}{1+Cu^\vv}
\label{e280}
\end{equation}
and
\begin{equation}
\lambda(\qq) = 1 + Cw(\qq)
\label{e290}
\end{equation}
with the same constant $C$ appearing in both equations.
Not unexpectedly the correspondence defines the waiting
time distribution uniquely up to a constant while the 
structure function is related to the Fourier transform
of the transition rates.

To invert the Laplace transformation in (\ref{e280})
and exhibit the form of the waiting time density $\psi(t)$
in the time domain one rewrites eq. (\ref{e280})
\begin{equation}
\psi(u) = \frac{u^{-\vv}}{C}\frac{1}{1+\frac{u^{-\vv}}{C}} =
\sum_{k=0}^\infty\frac{(-1)^k}{C^{k+1}}u^{-\vv(k+1)}
\end{equation}
and inverts the series term by term.
This yields the result
\begin{equation}
\psi(t;\vv,C) = \frac{t^{\vv-1}}{C}\sum_{k=0}^{\infty}
\frac{1}{\Gamma(\vv k+\vv)}\left(-\frac{t^\vv}{C}\right)^k .
\label{e2170}
\end{equation}
First one notes that for $\vv=1$ the result reduces to
\begin{equation}
\psi(t;1,C)=\frac{1}{C}\exp(-t/C)
\end{equation}
the familiar exponential form.
For $0<\vv<1$ the result is recognized as
\begin{equation}
\psi(t;\vv,C)=\frac{t^{\vv-1}}{C}E_{\vv,\vv}\left(-\frac{t^\vv}{C}\right) .
\label{e2190}
\end{equation}
where the function
\begin{equation}
E_{\alpha,\beta}(z)=\sum_{k=0}^\infty\frac{z^k}{\Gamma(\alpha k +\beta)}
\end{equation}
is the generalized Mittag-Leffler function \cite{erd55}.

The asymptotic behaviour of $\psi(t)$ for $t\rao$ is readily
obtained by noting that $E_{\vv,\vv}(0)=1$. 
Hence $\psi(t)$ behaves as
\begin{equation}
\psi(t)\propto t^{-1+\vv}
\label{e2180}
\end{equation}
for $t\rao$. 
For $0<\vv< 1$ the waiting time density becomes singular at the
origin. 

The asymptotic behaviour for large waiting times $t\rai$ 
is obtained from the asymptotic series expansion for the 
Mittag-Leffler function \cite{erd55}
\begin{equation}
E_{\alpha,\beta}(z)=
-\sum_{n=1}^N\frac{z^{-n}}{\Gamma(\beta-\alpha n)} + O(|z|^{-N})
\end{equation}
valid for $|$arg$(-z)|<(1-(\alpha/2))\pi|$ and $z\rai$.
It follows from this that $E_{\alpha,\beta}(-x)\propto x^{-2}$ 
for $x\rai$ and hence that
\begin{equation}
\psi(t)\propto t^{-1-\vv}
\label{e2200}
\end{equation}
for large $t\rai$ and $0<\vv<1$. This result shows that the waiting time 
distribution has an algebraic tail as it is usually assumed
in the theory of continuous time random walks 
\cite{shl74,tun74,tun75,SKW82,KBS87}.

In summary it has been shown that the master equation with a
Riemann-Liouville fractional time derivative of order $\vv$
corresponds exactly to a continuous time random walk whose
waiting time density is related to the generalized Mittag-Leffler
function and exhibits a power-law tail.
It would be interesting to know the precise conditions under which
the more general class of separable continuous time random walks obeying 
eq. (\ref{e2200}) ``approximates'' \cite{com96} this result.


\begin{thebibliography}{10}

\bibitem{hil93e}
R.~Hilfer, ``Classification theory for anequilibrium phase transitions,'' {\em
  Phys. Rev. E}, vol.~48, p.~2466, 1993.

\bibitem{hil95c}
R.~Hilfer, ``Fractional dynamics, irreversibility and ergodicity breaking,''
  {\em Chaos, Solitons {\&} Fractals}, vol.~5, p.~1475, 1995.

\bibitem{hil95e}
R.~Hilfer, ``Foundations of fractional dynamics,'' {\em Fractals}, vol.~3,
  p.~549, 1995.

\bibitem{hil95f}
R.~Hilfer, ``An extension of the dynamical foundation for the statistical
  equilibrium concept,'' {\em Physica A}, vol.~221, p.~89, 1995.

\bibitem{hil98c}
R.~Hilfer, {\em Applications of Fractional Calculus in Physics}.
\newblock Singapore: World Scientific Publ. Co., 1998.
\newblock , in Vorbereitung.

\bibitem{fox61}
C.~Fox, ``The {$G$} and {$H$} functions as symmetrical {F}ourier kernels,''
  {\em Trans. Am. Math. Soc.}, vol.~98, p.~395, 1961.

\bibitem{fel71}
W.~Feller, {\em An Introduction to Probability Theory and Its Applications},
  vol.~II.
\newblock New York: Wiley, 1971.

\bibitem{hil95a}
R.~Hilfer and L.~Anton, ``Fractional master equations and fractal time random
  walks,'' {\em Phys.Rev.E, Rapid Commun.}, vol.~51, p.~848, 1995.

\bibitem{hil95b}
R.~Hilfer, ``Exact solutions for a class of fractal time random walks,'' {\em
  Fractals}, vol.~3(1), p.~211, 1995.

\bibitem{KBS87}
J.~Klafter, A.~Blumen, and M.~Shlesinger, ``Stochastic pathway to anomalous
  diffusion,'' {\em Phys. Rev. A}, vol.~35, p.~3081, 1987.

\bibitem{com96}
A.~Compte, ``Stochastic foundations of fractional dynamics,'' {\em Phys.Rev.
  E}, vol.~55, p.~4191, 1996.

\bibitem{MKS98}
J.~K. R.~Metzler and I.~Sokolov.
\newblock preprint.

\bibitem{BN70}
M.~Barber and B.~Ninham, {\em Random and Restricted Walks}.
\newblock New York: Gordon and Breach Science Publ., 1970.

\bibitem{MW79}
E.~Montroll and B.~West, ``On an enriched collection of stochastic processes,''
  in {\em Fluctuation Phenomena} (E.~Montroll and J.~Lebowitz, eds.),
  (Amsterdam), p.~61, North Holland Publ. Co., 1979.

\bibitem{WR83}
G.~Weiss and R.~Rubin, ``Random walks: Theory and selected applications,'' {\em
  Adv. Chem. Phys.}, vol.~52, p.~363, 1983.

\bibitem{HK87}
J.~Haus and K.~Kehr, ``Diffusion in regular and disordered lattices,'' {\em
  Phys.Rep.}, vol.~150, p.~263, 1987.

\bibitem{hug95}
B.~Hughes, {\em Random Walks and Random Environments}, vol.~1.
\newblock Oxford: Clarendon Press, 1995.

\bibitem{hug96}
B.~Hughes, {\em Random Walks and Random Environments}, vol.~2.
\newblock Oxford: Clarendon Press, 1996.

\bibitem{erd55}
A.~{Erdelyi (et al.)}, {\em Higher Transcendental Functions}, vol.~III.
\newblock New York: Mc Graw Hill Book Co., 1955.

\bibitem{shl74}
M.~Shlesinger, ``Asymptotic solutions of continuous time random walks,'' {\em
  J. Stat. Phys.}, vol.~10, p.~421, 1974.

\bibitem{tun74}
J.~Tunaley, ``Asymptotic solutions of the continuous time random walk model of
  diffusion,'' {\em J. Stat. Phys.}, vol.~11, p.~397, 1974.

\bibitem{tun75}
J.~Tunaley, ``Some properties of the asymptotic solutions of the
  {M}ontroll-{W}eiss equation,'' {\em J. Stat. Phys.}, vol.~12, p.~1, 1975.

\bibitem{SKW82}
M.~Shlesinger, J.~Klafter, and Y.~Wong, ``Random walks with infinite spatial
  and temporal moments,'' {\em J. Stat. Phys.}, vol.~27, p.~499, 1982.

\end{thebibliography}
\end{document}